\documentclass[conference]{IEEEtran}

\usepackage{amsfonts}
\usepackage{stackengine}
\usepackage{cleveref}

\usepackage{amsmath}
\usepackage{graphicx}
\usepackage{lipsum}
\usepackage{cite}

\begin{document}

\title{Fault tolerant controller design for a class of affine nonlinear systems based on adaptive virtual actuator }

\author{S. Narges~Mahdian Zadeh, Reza~Ghasemi
	\thanks{S.N. Mahdian Zadeh with MSc student of Electrical Engineering, University of Qom,
	Qom,Iran, e-mail: n.mahdianzadeh@stu.qom.ac.ir}	 
	
	\thanks{R. Ghasemi  with the Department
		of Electrical Engineering, University of Qom, Qom,
		Iran, e-mail: r.ghasemi@qom.ac.ir}}


\maketitle

\begin{abstract}
	
An Adaptive Fault-tolerant Controller procedure for a class of the affine nonlinear system is developed in this paper. This methodology hides both the faults and external disturbances. Compare to the procedure that require separate fault detection, isolation, and identification units, the suggested method concentrates on fault hiding to reduce the units. The main merits of the proposed method are 1) guaranteed stability of the closed loop system, 2) convergence of the states of nominal and faulty systems to zero, and 3) convergence of the tracking error to the origin. The robustness of presented method is assured against external disturbances. Simulation results illustrate the effectiveness of proposed fault tolerant control.         	
		
\end{abstract}

\begin{IEEEkeywords}
 fault tolerant control, adaptive virtual actuator,fault hiding.
\end{IEEEkeywords}

\section{Introduction}
The demand for more productivity  has challenged   many modern industrial systems.  Such conditions increase the probability of system faults and change in system dynamics\cite{1}. A fault is described as an unpermitted deflection of at least one characteristic feature or parameter of the system from the standard condition\cite{2}. A fault usually involves a sensor or an actuator, which results in the loss of their correct performance. In order to improve the reliability of a system, the Fault Tolerant Control(FTC) is expressed. A fault-tolerant control system is automatically compensating the effects of faults which is kept the acceptable  performance of the closed loop system\cite{1}. Actuallyَ, the main aim of FTC approaches is to preserve the demanded performance of the system or at  least  its  stability  when the faults occur\cite{3}. The  FTC design techniques mainly can be classified to passive or the active approach\cite{4}. In the passive FTC techniques, the control laws  take into account the fault as a system perturbation \cite{5}. Thus, they have a fixed parameter  in order to keep the performance of the closed-loop system in an acceptable level when the presence known structure faults\cite{3}. For example in FTC research, Fekih and Pilla designed a passive FTC approach for an F-18 aircraft model. They used a sliding surface to eliminate the faults \cite{6}. In active FTC approaches, parameters and structure of the nominal controller are reconfigured according to the occurred fault\cite{3}. In this cases, a fault diagnosis, detection and isolation (FDI)unit is used. The model-based is the example of this unit's design. But with using a model-based approach, there will be a difference between the actual system and its mathematical model\cite{5}. Another class of active FTC approaches are based on an adaptive control which doesn't require an FDI unit\cite{7}. A successful example in this design, is fault hiding method. This block that hides the fault, is adding  between the controller and the system. This method doesn't need to the readjusting of the controller\cite{5} . When faults of sensors and actuators occur, the virtual sensors and actuators are well known fault hiding blocks.\cite{8}. In \cite{9}, the fault tolerant control (FTC) is investigated for linear systems using the fault-hiding approach. Fault hiding approach is used for nonlinear systems,too. In\cite{10}, virtual actuator synthesis is presented for Lipschitz non-linear systems.

 In \cite{11}, is studied a dual-timescale aircraft flight control problem with using the Sliding Mode Control (SMC). SMC is the well-known robust control technique, because  has been  used to improve the stability and performance   in the presence of uncertainties\cite{2}. In \cite{12}, a sliding  mode-based FTC is designed for the Boeing 747 that is a civilian fixed-wing aircraft. Also, in\cite{13}, a passive actuator FTC is investigated for the MIMO affine class of nonlinear systems with using Sliding Mode Control.\\
For identifying unknown nonlinear characteristics, neural networks and fuzzy logic systems are effective,too\cite{14}. For example, an adaptive fuzzy control method for strict-feedback systems is provided in\cite{15}.\cite{16} has extended the results in \cite{15}, with using an adaptive neural network controller for nonlinear systems. Also, in\cite{17}, for systems with unknown control directions and unmodeled dynamics is proposed an adaptive fault-tolerant controller  which is  based on fuzzy approximation.
 In \cite{18} is designed a neural networks-based adaptive Finite-time  fault-tolerant  control for  a  class  of  switched  nonlinear systems in lower-triangular  form  under  arbitrary  switching  signals. A neural-adaptive output-feedback controller is designed for apply to transportation vehicles based on wheeled inverted pendulum in \cite{19},too. Also, the observer-based adaptive neural fault-tolerant tracking control for nonlinear systems in non-strict feedback form has been presented in\cite{14}. Another example of a controller design in\cite{20} is presented which an adaptive neural network(NN)-based fault-tolerant control is  studied  for the compensation of actuator failures in nonlinear systems with time-varying delay. 
 
Compare  to the researches which concentrate on system with constant input gain, this paper deals with the system with nonlinear input gain. the virtual adaptive actuator design is developed for a class on affine nonlinear system in presence of the actuator fault.
The most important goals of the proposed methodology are as:
1) guaranteed stability of the closed loop system, 2) promising performance of fault hiding in the presence of disturbance and uncertain dynamic, 3)  convergence of tracking error to zero, 4) convergence of the error between health and the faulty system.
 
The remainder of the paper is organized as follows. The section 2 presents problem formulation and required theories. The Adaptive virtual actuator is derived in section 3. Then, section 4 states the results of the simulation. Finally, this paper ends with conclusion in Section 5.  

\section{ Problem Formulation}
Consider the following nominal plant:

\begin{equation} \label{eq:1}
\left\{ \begin{matrix}
\dot{\hat{x}}(t)=A\hat{x}(t)+b[f(\hat{x}(t))+g(\hat{x}(t))u(t)]  \\
\hat{y}=C\hat{x}(t).  \\
\end{matrix} \right.
\end{equation}\\

where $\hat{x} \in \mathbb{R}^{n} $ is the state vector, $u(t) \in \mathbb{R}^{m}$ is the control input and $\hat{y} \in \mathbb{R}^{l}$is the output vector. $A\in {{\mathbb{R}}^{n\times n}},b\in {{\mathbb{R}}^{n\times 1}}$ and $C\in {{\mathbb{R}}^{l\times n}}$ are known constant matrices such that the system is controllable. $f(.):{{\mathbb{R}}^{n}}\to {{\mathbb{R}}}$ is a Lipschitz function with respect to ${\hat{x}}(t)$ and $g(.):{{\mathbb{R}}^{n}}\to {{\mathbb{R}}}$ is continuously derivative; so $ g(0) \ne 0$.

One of the goals of this paper is that  the states of  nominal system track the states of desire system. The dynamic of desire plant is as follows:

\begin{equation} \label{eq:2}
{{\dot{x}}_{d}}(t)=A_{d}{{x}_{d}}(t)+B_{d}r(t),
\end{equation}\\
where $A_{d}\in {{\mathbb{R}}^{n\times n}},B_{d}\in {{\mathbb{R}}^{n\times 1}}$  are known constant matrices, $x_{d}(t) \in \mathbb{R}^{n} $ is the  desire state vector and $r(t)$ is the reference input.

So,the nominal controller for the nominal plant is given as equation(3) until This  closed-loop system  is stable.\\

\begin{equation} \label{eq:3}
u(t) = \frac{1}{{g(\hat x(t))}}[ - f(\hat x(t)) + {k_1}r(t) + {k_2}\hat x(t)],
\end{equation}\\
where $k_{1}$ and $k_{2}$ is as follows:
\begin{equation} \label{eq:4}
{k_1} = {({b^T}b)^{ - 1}}{b^T}({A_d} - A),
\end{equation}\\
\begin{equation} \label{eq:5}
{k_2} = {({b^T}b)^{ - 1}}{b^T}{B_d}.
\end{equation}\\

\textit{\textbf{Theorem1}}:Consider the dynamic of the nominal system in equations(1). The proposed nominal control input in equations(3) through (5), makes this system asymptotically stable and $e(t)$ achieve to the zero, if there exist a matrix the positive definite matrix ${{p}_{1}}\in {{\mathbb{R}}^{n\times n}}$ which is satisfied in the following  inequalities:
\begin{equation} \label{eq:6}
-{{Q}_{1}}={{A}^{T}}{{p}_{1}}+{{p}_{1}}A<0.
\end{equation}\\

\textbf{\textit{proof:}}
The difference between the states of nominal and desire is as(7):
\begin{equation} \label{eq:7}
e(t) = \hat x(t) - {x_d}(t).
\end{equation}\\
 
The time derivative of this error is as follows:
\begin{equation} \label{eq:8}
\begin{array}{l}
\dot e(t) = A\hat x(t) + bf\left( {\hat x(t)} \right) + bg\left( {\hat x(t)} \right)u\\\\

- {A_d}{x_d}(t) - {B_d}r(t).
\end{array}
\end{equation}\\

That according to the definition of error, the following equation holds.
\begin{equation} \label{eq:9}
\begin{array}{l}
\dot e(t) = A\hat x(t) + bf\left( {\hat x(t)} \right) + bg\left( {\hat x(t)} \right)u \\\\

 - {A_d}(\hat x(t) - e(t)) - {B_d}r(t).
\end{array}
\end{equation}\\

Using equation (3), the above can be rewritten as:
\begin{equation} \label{eq:10}
\dot e(t) = (A - {A_d} + b{k_2})\hat x(t) + {A_d}e(t) + (b{k_1} - {B_d})r.
\end{equation}\\

Then, the following lyapunov function is candidate as below.
\begin{equation} \label{eq:11}
V_{1}(t) = \frac{1}{2}{e^T}(t){p_1}e(t),
\end{equation}\\
where ${p_1} \in {\mathbb{R}^{n \times n}}$ is a constant matrix.\\ 
 
The time derivative of the equation (11) is as:
\begin{equation} \label{eq:12}
\begin{array}{l}
\dot V_{1}(t) = \frac{1}{2}e{(t)^T}{p_1}\dot e(t) + \frac{1}{2}\dot e{(t)^T}{p_1}e(t), 
\end{array}
\end{equation}\\
Based on the equation(10), we have:
\begin{equation} \label{eq:13}
\begin{array}{*{20}{l}}
\begin{array}{l}
{{\dot V}_1}(t) = \frac{1}{2}e{(t)^T}({A_d}^T{p_1} + {p_1}{A_d})e(t) + \frac{1}{2}{e^T}(t){p_1}A\hat x(t)\\
{}
\end{array}\\
\begin{array}{l}
- \frac{1}{2}{e^T}(t){p_1}{A_d}\hat x(t) + \frac{1}{2}{e^T}(t){p_1}bg(\hat x(t)){k_2}\hat x(t)\\
{}
\end{array}\\
\begin{array}{l}
+ \frac{1}{2}{{\hat x}^T}(t){A^T}{p_1}e(t) - \frac{1}{2}{{\hat x}^T}(t){A_d}^T{p_1}e(t) + \\
{}
\end{array}\\
{ + \frac{1}{2}{{\hat x}^T}(t){k_2}^T{g^T}(\hat x(t)){b^T}{p_1}e(t) + {e^T}(t){p_1}bg(\hat x(t)){k_1}r}\\\\

{ - {e^T}(t){p_1}{B_b}r(t).}
\end{array}
\end{equation}\\

The above equation can be arranged as follows:
\begin{equation} \label{eq:14}
\begin{array}{*{20}{l}}
\begin{array}{l}
{{\dot V}_1}(t) =  - \frac{1}{2}{e^T}(t){Q_1}e(t) + \\
{}
\end{array}\\
\begin{array}{l}
\frac{1}{2}{e^T}(t){p_1}(A - {A_d} + bg(\hat x(t)){k_2})\hat x(t) + \\
{}
\end{array}\\
\begin{array}{l}
\frac{1}{2}{x^T}(t)(k_2^T{g^T}(\hat x(t)){b^T} + {A^T} - A_d^T){p_1}e(t)\\
{}
\end{array}\\
{ + {e^T}(t){p_1}(bg(\hat x(t)){k_1} - {B_d})r(t).}
\end{array}
\end{equation}\\

Using (4) and (5), the equation(14) is rewritten as (15)
\begin{equation} \label{eq:15}
\dot V_{1}(t) =  - \frac{1}{2}{e^T}(t){Q_1}e(t) < 0.
\end{equation}\\

So, this theory proves the asymptotically stability of the closed loop system. 

Now, consider the following faulty plant:
\begin{equation} \label{eq:16}
\left\{ {\begin{array}{*{20}{c}}
	\begin{array}{l}
	{{\dot x}_f}(t) = A{x_f}(t) + bf({x_f}(t)) + \\
	{b_f}[g({x_f}(t))({u_f}(t) + {d_f}(t))] + \\
	Ed(t)
	\end{array}\\
	{}\\
	{{y_f}(t) = C{x_f}(t),}
	\end{array}} \right.
\end{equation}\\
where ${{x}_{f}}(t)\in {{\mathbb{R}}^{n}}$ ,${{x}_{f}}(0)={{x}_{0}}$,  is  the  state  of  faulty  plant, ${{u}_{f}}(t)\in {{\mathbb{R}}^{m}}$ is the control input and ${{y}_{f}}(t)\in {{\mathbb{R}}^{l}}$ is the faulty plant output. ${{\left\| {d(t)} \right\|}_{2}}\le {{d}_{\max }}$  and ${{\left\| \dot{d}(t) \right\|}_{2}}\le {{\dot{d}}_{\max }}$ ,$t\ge 0 $ where $d_{max}$  and ${{\dot{d}}_{\max }}$ are constants and $E\in {{\mathbb{R}}^{n\times q}}$ is known constant matrix. It should be emphasized that ${{u}_{f}}(t)$  is  used  to  distinguish  the  control  input  in  the faulty plants and the loss of effectiveness actuator fault is formulated as the change in the input matrix b as follows\cite{3}:\\
\begin{center}
	${{b}_{f}}=b\Theta \,,\,\,\,\,\,\,\,\,\,\,\,\,\,\,\,\,\,\Theta \overset{\Delta }{\mathop{=}}\,\,diag({{\theta }_{1}},{{\theta }_{2}},...,{{\theta }_{m}})$
	
\end{center}
where ${{\theta }_{i}}\,,\,\,i=1,...,m$ is unknown constants such that $0<{{\theta }_{i}}\le 1$ and ${{\theta }_{i}}=1$ is defined for the healthy actuator.\\

The following theories and lemma will be used in next section.\\

\textbf{\textit{Theorem2\cite{21}:}} Assume that $f:{R^n} \times {R^m} \to {R^n}$ is continuously differentiable at each point $(x,y)$ of an open set $S \subset {R^n} \times {R^m}$. Let $({x_0},{y_0})$ be a point in $S$ for which $f({x_0},{y_0}) = 0$ and for which the Jacobian matrix $\left[ {\frac{{\partial f}}{{\partial x}}} \right]({x_0},{y_0})$ is nonsingular.Then there exist neighborhoods $U \subset {R^n}$ of ${x_0}$ and $V \subset {R^m}$ of ${y_0}$ such that for each $y \in V$ the equation $f(x,y) = 0$ has a unique solution $x \in U$. Moreover, this solution can be given as $x = g(y)$, where $g$ is continuously differentiable at $y = {y_0}$. 
This theory has been proven in \cite{21}.

\textbf{\textit{Theorem3\cite{21}:}}Assume that $f:{R^n} \times {R^m} \to {R^n}$ is continuously differentiable at each point $x$ of an open set $S \subset {R^n}$. Let $x$ and $y$  be two points of $S$ such that the line segment $L(x,y) \subset S$. Then there exists a point $z$ of $L(x,y)$ such that  
\begin{center}
$f(y) - f(x) = \frac{{\partial f}}{{\partial x}}\left| {_{x = z}\,(y - x)} \right.$
\end{center}        
The proof of this theory is mentioned in \cite{21}.\\

The "Lemma 1" is result from  the Theory 3, too.

\textbf{\textit{Lemma1\cite{21}:}} Let $f:\,\left[ {a,b} \right] \times D \to {R^m}$ be continuous for some domain $D \subset {R^n}$. Suppose that $\frac{{\partial f}}{{\partial x}}$ exists and is continuous on $\left[ {a,b} \right] \times D$. If, for a convex subset 
$V \subset D$, there is a constant $L \ge 0$ such that  
\begin{center}
$\left\| {\frac{{\partial f}}{{\partial x}}(t,x)} \right\| \le L$
\end{center}
on $\left[ {a,b} \right] \times V$ then 

\begin{center}
$\left\| {f(t,x) - f(t,y)} \right\| \le L\left\| {x - y} \right\|$
\end{center}
for all $t \in \left[ {a,b} \right]\,,\,x \in V\,,\,and\,y \in V$.

\textbf{\textit{Theorem4\cite{21}:}}
Let $D \subset {R^n}$ be a domain that contains the origin and $V:\left[ {0,\infty } \right) \times D \to R$ be a continuously differentiable function sush that 
\begin{center}
	${\alpha _1}(\left\| x \right\|) \le V(t,x) \le {\alpha _2}(\left\| x \right\|)$
\end{center}
\begin{center}
	$\frac{{\partial V}}{{\partial t}} + \frac{{\partial V}}{{\partial x}}f(t,x) \le  - {W_3}(x)\,\,,\,\,\,\forall \left\| x \right\| \ge \mu  > 0$
\end{center}
$\,\forall t \ge 0$ and $\,\forall x \in D$, where ${\alpha _1}$ and ${\alpha _2}$ are class $\mathcal{K}$ functions and ${W_3}(x)$ is a continuous positive definite function. Take $r > 0$ such that ${B_r} \subset D$ and suppose that
\begin{center}
	$\mu  < \alpha _2^{ - 1}({\alpha _1}(r))$
\end{center}
then, there exists a class $\mathcal{KL}$ function $\beta $ and for every initial state $x({t_0})$, satisfying $\left\| {x({t_0})} \right\| \le \alpha _2^{ - 1}({\alpha _1}(r))$, there is$T \ge 0$(dependent on $x({t_0})$ and $\mu $) such that the solution of $\dot x(t) = f(t,x)$ satisfies
\begin{center}
	$\left\| {x(t)} \right\| \le \beta (\left\| {x({t_0})} \right\|,t - {t_0}),\,\,\,\forall {t_0} \le t \le {t_0} + T$
\end{center}
\begin{center}
	$\left\| {x(t)} \right\| \le \alpha _1^{ - 1}({\alpha _2}(\mu )),\,\,\,\forall t \ge {t_0} + T$,
\end{center}
moreover, if $D = {R^n}$ and ${\alpha _1}$ belongs to class ${\mathcal{K}_\infty }$, then the above inequalities hold for any initial state $x({t_0})$, with no restriction on how large $\mu $ is.
This theory has been proven in \cite{21}.

\section{Adaptive Virtual Actuator Design }
In this section, an adaptive virtual actuator controller is designed. It can recover the performance of the close-loop system after the occurrence of actuator faults and external disturbancees. In this method, the applied fault-hiding approach does not need to change the nominal controller when the faults have occurred.\\

The dynamic of the difference system between the nominal and faulty systems is definded as follows:\\
\begin{equation} \label{eq:17} 
\left\{ {\begin{array}{*{20}{c}}
	{\dot { \tilde{x}}(t) = A\tilde x(t) + bf({x_f}(t)) + {b_f}g({x_f}(t)){u_f}(t)}\\
	\begin{array}{l}
	+ {b_f}g({x_f}(t)){d_f}(t) - bf(\hat x(t)) - bg(\hat x(t))u(t)\\
	+ Ed(t)
	\end{array}\\
	{}\\
	{\tilde y = C\tilde x(t),}
	\end{array}} \right.
\end{equation}\\
where $\tilde{x}(t)\overset{\Delta }{\mathop{=}}\,{{x}_{f}}(t)-\hat{x}(t)$ and $\tilde{y}(t)\overset{\Delta }{\mathop{=}}\,{{y}_{f}}(t)-\hat{y}(t)$.\\

\textit{Proposition1:}Based on theory 2, there exist an ideal controller as $u_{f}^{*}(t)$ in which satisfy the following equation.
\begin{equation} \label{eq:18}
bf({{x}_{f}}(t))+bg({{x}_{f}}(t))u_{f}^{*}(t)=0
\end{equation}\\

In this approach, there exist ${M^*}(t) \in {\mathbb{R}^{m \times n}}\,$ and ${N^*}(t) \in {\mathbb{R}^{m \times m}}$ such that $u_{f}^{*}(t)$ is defined as:
\begin{equation} \label{eq:19}
u_{f}^{*}(t)={{M}^{{{*}^{T}}}}(t)\tilde{x}(t)+{{N}^{{{*}^{T}}}}(t)u(t).
\end{equation}\\ 

The faulty plant control input ${{u}_{f}}(t)$, is considered to be an estimation of ${{u}_{f}^{*}}(t)$,
\begin{equation} \label{eq:20}
{{u}_{f}}(t)=M(t)\tilde{x}(t)+N(t)u(t)-\hat{d}(t),
\end{equation}\\
where $M(t)\in {{\mathbb{R}}^{m\times n}},N(t)\in {{\mathbb{R}}^{m\times m}}$ and $\hat{d}(t)\in {{\mathbb{R}}^{m}}$ are obtained from:

\begin{equation} \label{eq:21}
 \dot{M}(t)=-{{\gamma }_{1}}{{\tilde{x}}^{T}}(t){{g}^{T}}({{x}_{f}}(t)){{b}^{T}}{{p}_{1}}\tilde{x}(t)
\end{equation}\\
\begin{equation} \label{eq:22}
\dot{N}(t)=-{{\gamma }_{2}}{{u}^{T}}(t){{g}^{T}}({{x}_{f}}(t)){{b}^{T}}{{p}_{1}}\tilde{x}(t)
\end{equation}\\
\begin{equation} \label{eq:23}
{{\dot{\hat{d}}}}(t)={{\gamma }_{3}}{{g}^{T}}({{x}_{f}}(t)){{b}^{T}}{{p}_{1}}\tilde{x}(t)
\end{equation}\\
where ${{\gamma }_{1}}\in {{\mathbb{R}}^{n\times n}}$ and ${{\gamma }_{2}}\in {{\mathbb{R}}^{m\times m}}$ are positive definite gain matrices. ${{\gamma }_{3}}$ is a positive constant and $p_1$ is a positive definite matrix to be designed.

\textit{\textbf{Theorem5}}: Consider the dynamic of reconfigured closed-loop system delivered in equations (3) and (16). The proposed faulty plant control input in equation (20) and update laws(21) through (23), makes this system uniformly ultimately bounded, if there exist a positive definite matrix ${{p}_{2}}\in {{\mathbb{R}}^{n\times n}}$ which is satisfied in the following  inequalities:

\begin{equation} \label{eq:24}
-{{Q}_{2}}={{A}^{T}}{{p}_{2}}+{{p}_{2}}A<0
\end{equation}\\

\textit{\textbf{Proof}}:
The difference between $u_{f}^{*}(t)$ and $u_{f}(t)$ is as follows:
\begin{equation} \label{eq:25}
{{e}_{u}}=u_{f}^{*}(t)-u_{f}(t)={{\tilde{M}}^{T}}(t)\tilde{x}(t)+{{\tilde{N}}^{T}}(t)u(t)+\hat{d}(t) 
\end{equation}\\
where $\tilde{M}(t)={{M}^{*}}(t)-M(t)$ and $\tilde{N}(t)={{N}^{*}}(t)-N(t)$.

Paying attention to that there exists ${{Q}^{*}}\in {{\mathbb{R}}^{m\times q}}$ such that $b_{f}g({{x}_{f}}(t)){{Q}^{*}}=E$, the equation(17)  is rewritten as (26):
\begin{equation} \label{eq:26}
\begin{array}{*{20}{l}}
\begin{array}{l}
\dot {\tilde {x}}(t) = A\tilde x(t) + bf({x_f}(t)) + {b_f}g({x_f}(t)){u_f}(t)\\
{}
\end{array}\\
\begin{array}{l}
+ {b_f}g({x_f}(t)){d_f}(t) + {b_f}g({x_f}(t)){Q^*}d(t) - bf(\hat x(t))\\
{}
\end{array}\\
{ - bg(\hat x(t))u(t).}
\end{array}
\end{equation}\\

Using ${{d}_{f}}(t)+{{Q}^{*}}d(t)={{d}_{t}}(t)$, the equation (26) is rewritten as follows.
\begin{equation} \label{eq:27}
\begin{array}{*{20}{l}}
\begin{array}{l}
\dot {\tilde{ x}}(t) = A\tilde x(t) + bf({x_f}(t)) + {b_f}g({x_f}(t)){d_t}\\
{}
\end{array}\\
\begin{array}{l}
+ {b_f}g({x_f}(t))[{u_f}(t) - u_f^*(t) + u_f^*(t)] - bf(\hat x(t))\\
{}
\end{array}\\
\begin{array}{l}
- bg(\hat x(t))u(t) = A\tilde x(t) - {b_f}g({x_f}(t)){e_u}(t) + \\
{}
\end{array}\\
{{b_f}g({x_f}(t)){d_t} - bf(\hat x(t)) - bg(\hat x(t))u(t).}
\end{array}
\end{equation}\\

 According to the equation (18), we have:
 \begin{equation} \label{eq:28}
\begin{array}{*{20}{l}}
 \begin{array}{l}
 \dot {\tilde{ x}}(t) = A\tilde x(t) - {b_f}g({x_f}(t)){e_u}(t) + \\
 {}
 \end{array}\\
 {{b_f}g({x_f}(t)){d_t} - bf(\hat x(t)) - bg(\hat x(t))u(t).}
 \end{array}
 \end{equation}\\
 
Also, with using(25), the following equation is obtained:
\begin{equation} \label{eq:29}
\begin{array}{*{20}{l}}
\begin{array}{l}
\dot{ \tilde{ x}}(t) = A\tilde x(t) - {b_f}g({x_f}(t)){{\tilde M}^T}(t)\tilde x(t) - \\
{}
\end{array}\\
\begin{array}{l}
{b_f}g({x_f}(t)){{\tilde N}^T}(t)u(t) - {b_f}g({x_f}(t))\hat d(t) + \\
{}
\end{array}\\
{{b_f}g({x_f}(t)){d_t} - bf(\hat x(t)) - bg(\hat x(t))u(t).}
\end{array}
\end{equation}\\

Then $\tilde{d}(t)={{d}_{t}}(t)-\hat{d}(t)$  is defined and the  above equation is rewritten as follows.
\begin{align}\label{eq:30}\nonumber  
& \dot{\tilde{x}}(t)=A\tilde{x}(t)+b_{f}g({{x}_{f}}(t))\tilde{d}(t)-bf(\hat{x}(t))-bg(\hat{x}(t))u(t)- \\\nonumber 
& b_{f}g({{x}_{f}}(t)){{{\tilde{M}}}^{T}}\tilde{x}(t)-b_{f}g({{x}_{f}}(t)){{{\tilde{N}}}^{T}}u(t). \\ 
\end{align}

To prove the stability of the closed-loop system, the following lyapunov function is candidate.
\begin{equation} \label{eq:31}
\begin{array}{l}
{V_2}(t) = \frac{1}{2}{{\tilde x}^T}(t){p_2}\tilde x(t) + \frac{1}{2}\frac{{{{\tilde d}^2}(t)}}{{{\gamma _1}}} + \\
\\
\frac{1}{2}\frac{{{{\tilde M}^T}(t)\tilde M(t)}}{{{\gamma _2}}} + \frac{1}{2}\frac{{{{\tilde N}^T}(t)\tilde N(t)}}{{{\gamma _3}}}
\end{array}
\end{equation}
where ${{\gamma }_{1}},{{\gamma }_{2}},{{\gamma }_{3}}$ are scalar values.\\

The time derivative of the equation(31) is as below.
\begin{equation} \label{eq:32}
\begin{array}{*{20}{l}}
\begin{array}{l}
{{\dot V}_2}(t) = \frac{1}{2}{{\dot{ \tilde{ x}}}^T}(t){p_2}\tilde x(t) + \frac{1}{2}{{\tilde x}^T}(t){p_2}\dot {\tilde {x}}(t) + \frac{{\tilde d(t)\dot{ \tilde {d}}(t)}}{{{\gamma _1}}}\\
{}
\end{array}\\
{ + \frac{{{{\tilde M}^T}(t){{\dot {\tilde {M}}}^T}(t)}}{{{\gamma _2}}} + \frac{{{{\tilde N}^T}(t)\dot{ \tilde {N}}(t)}}{{{\gamma _3}}},}
\end{array}
\end{equation}
that based on (30),we have:
\begin{equation} \label{eq:33}
\begin{array}{*{20}{l}}
\begin{array}{l}
{{\dot V}_2}(t) = \frac{1}{2}{{\tilde x}^T}(t)({A^T}{p_2} + {p_2}A)\tilde x(t) -\\
\\
 {{\tilde x}^T}(t){p_2}{b_f}g({x_f}(t)){{\tilde M}^T}(t)\tilde x(t)\\
{}
\end{array}\\
\begin{array}{l}
- {{\tilde x}^T}(t){p_2}{b_f}g({x_f}(t)){{\tilde N}^T}(t)u(t) +\\
\\
 {{\tilde x}^T}(t){p_2}{b_f}g({x_f}(t))\tilde d(t) - \\
{}
\end{array}\\
\begin{array}{l}
{{\tilde x}^T}(t){p_2}bf(\hat x(t)) - {{\tilde x}^T}(t){p_2}bg(\hat x(t))u(t) + \frac{{\tilde d(t)\dot {\tilde {d}}(t)}}{{{\gamma _1}}}\\
{}
\end{array}\\
{ - \frac{{{{\tilde M}^T}(t)\dot M(t)}}{{{\gamma _2}}} - \frac{{{{\tilde N}^T}(t)\dot N(t)}}{{{\gamma _3}}}}
\end{array}
\end{equation}
 
 Using (3), (21) through (23), the equation (33) is expressed as follows.
\begin{equation}\label{eq:34}
\begin{array}{*{20}{l}}
{\dot V_{2}(t) =  - \frac{1}{2}{{\tilde x}^T}(t){Q_2}\tilde x(t) - {{\tilde x}^T}(t){p_2}b{k_1}r - {{\tilde x}^T}(t){p_2}b{k_2}\hat x(t)}\\\\

{\begin{array}{*{20}{l}}
	{ + \frac{{\tilde d(t){{\dot d}_t}(t)}}{{{\gamma _2}}}}\\
	{}
	\end{array}}\\

\end{array}
\end{equation}

After some mathematical manipulations, the following inequality is obtained.
\begin{equation}\label{eq:35}
\begin{array}{*{20}{l}}
\begin{array}{l}
{{\dot V}_2}(t) \le  - \frac{1}{2}{{\tilde x}^T}(t){Q_2}\tilde x(t) + \left\| {\tilde x(t)} \right\|\left\| {{p_2}b{k_1}r} \right\| + \\
{}
\end{array}\\
{\left\| {\tilde x(t)} \right\|\left\| {{p_2}b{k_2}\hat x(t)} \right\| + \frac{{{{\tilde d}_{\max }}(t){{\dot d}_{t\max }}(t)}}{{{\gamma _2}}} + \theta {{\left\| {\tilde x(t)} \right\|}^2} - \theta {{\left\| {\tilde x(t)} \right\|}^2},}
\end{array}
\end{equation}\\
that for $\left\| {\tilde x(t)} \right\| > \frac{1}{{2\theta }}(\beta  + \sqrt {{\beta ^2} - 4\theta \mu }$ the following inequality is established, where $0 < \theta  < 1\,\,,\,\mu  = \frac{{{{\tilde d}_{\max }}{{\dot d}_{t\max }}}}{{{\gamma _2}}}$ and $\beta  = \left\| {{p_2}b{k_1}r} \right\| + \left\| {{p_2}b{k_2}\hat x} \right\|$ are scalar values and according to theory 1, $ \left\| {\hat x} \right\| < \varepsilon$.
 \\
 \begin{equation}\label{eq:36}
\dot V_{2}(t) \le \frac{1}{2}{\tilde x^T}(t)[ - {Q_2} + \theta ]\tilde x(t) + \beta \left\| {\tilde x(t)} \right\| - \theta {\left\| {\tilde x(t)} \right\|^2} + \mu  < 0
 \end{equation}
 
 So, the uniformly ultimately bounded of the closed loop system is achieved. 
 
 \begin{figure}
 	\centering 
 	\includegraphics[scale=0.25]{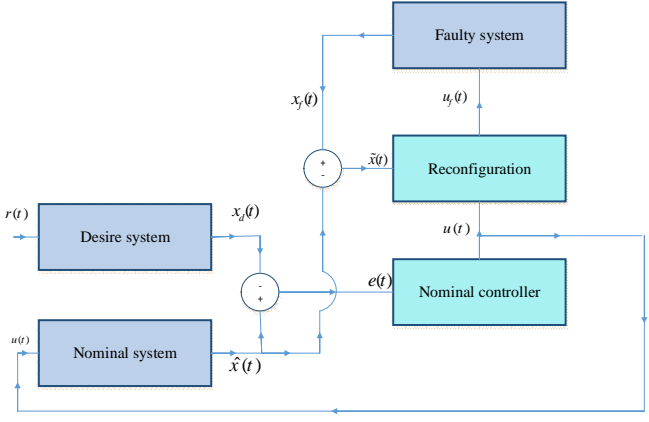}
 	\caption{The block diagram FTC with the virtual actuator.}
 	\label{Blockdiagram}
 \end{figure}

\section{Simulation result }
In this paper, the dynamics of the simulated system is considered as follows. 

\begin{equation}\label{eq:37}
\begin{array}{l}
\dot x(t) = \left[ {\begin{array}{*{20}{c}}
	0&1&0\\
	0&0&1\\
	{ - 1}&{ - 2}&{ - 3}
	\end{array}} \right]x(t) + \left[ {\begin{array}{*{20}{c}}
	0\\
	0\\
	1
	\end{array}} \right][f(x(t)) + \\\\

g(x(t))u(t)]+ Ed(t)\\
\\
y(t) = [\begin{array}{*{20}{c}}
1&1&{1]x(t)}
\end{array}
\end{array}
\end{equation}\\
 where $g(x(t))=0.5sin(t)+4\,\,,\,f(x(t))=0.05\sin ({{x}_{3}}(t))\,$ and $E=0.5bg(x(t))\,\,\,$ .Then  a loss of effectiveness actuator fault with $\Theta =0.65{{I}_{3}}$ occurs at ${{t}_{f}}=15s$. A disturbance applied in ${{t}_{d}}=20s$ and a time-varying additive actuator fault occurs at ${{t}_{fd}}=25s$.\\
 
 To design the proposed virtual actuator given by(21) through (23),we set${{\gamma }_{1}}=20\,,\,\,{{\gamma }_{2}}=200\,,\,\,{{\gamma }_{3}}=1000$ . Also,$\theta  = 0.5$.\\
 
The dynamics of desire agent is considered as follows.
 
 \begin{equation}\label{eq:38}
{\dot x_d}(t) = \left[ {\begin{array}{*{20}{c}}
 	0&1&0\\
 	0&0&1\\
 	{ - 1}&{ - 2}&{ - 4}
 	\end{array}} \right]{x_d}(t) + \left[ {\begin{array}{*{20}{c}}
 	0\\
 	0\\
 	1
 	\end{array}} \right]r(t),
 \end{equation}\\
where $r(t)$ is a reference input  and it can be the step function. Also, matrices ${{p}_{1}}$ and ${{p}_{2}}$ as
\begin{equation}\label{eq:39}
{{p}_{1}}= \begin{bmatrix}
2.5000 & 2.5000 & 0.5000 \\
2.5000 & 6.5000 & 1.5000 \\
0.5000 & 1.5000 & 0.5000 
\end{bmatrix},
\end{equation}

\begin{equation}\label{eq:40}
{{p}_{2}}= \begin{bmatrix}
2.8000 & 2.6000 & 0.5000 \\
2.6000 & 7.1000 & 1.8000 \\
0.5000 & 1.8000 & 1.1000 
\end{bmatrix}.
\end{equation}

Fig. 2  shows difference between the states of desire and nominal system, and consequently the nominal system track well the desire system. So, the nominal controller is worked correctly and $e$ achieve to zero, too.

\begin{figure}[!ht]
	\centering
	\includegraphics[scale=0.25]{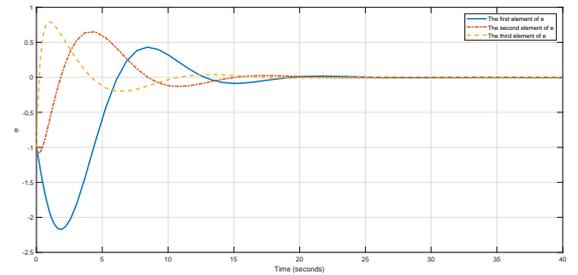}
	\caption{The error between states of the desire and nominal system.}
	\label{Blockdiagram}
\end{figure}
 
The output of nominal close-loop system without using the adaptive virtual actuator and desire system output are shown in Fig. 3.

 \begin{figure}[!ht]
	\centering
	\includegraphics[scale=0.25]{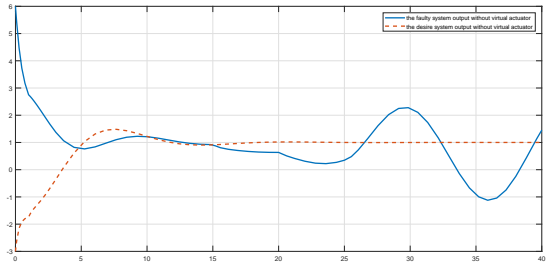}
	\caption{The output of close-loop system without using the adaptive virtual actuator.}
	\label{Blockdiagram}
\end{figure}

 Now, the output with using the proposed adaptive virtual actuator is shown in Fig. 4.  
 \begin{figure}[!ht]
	\centering
	\includegraphics[scale=0.25]{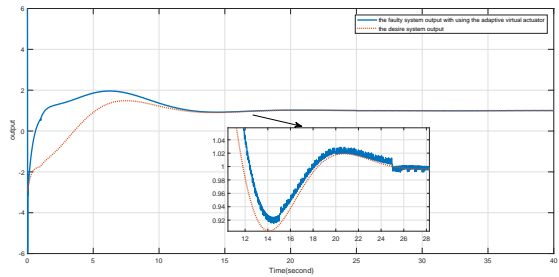}
	\caption{The output of close-loop system with using the adaptive virtual actuator.}
	\label{Blockdiagram}
\end{figure}

As shown in Fig. 4, after the fault occurred, the output tracks the desire one. \\
Fig. 5, Fig. 6 and Fig. 7 show the states of  faulty and  nominal system, too. \\

\begin{figure}[!ht]
	\centering
	\includegraphics[scale=0.25]{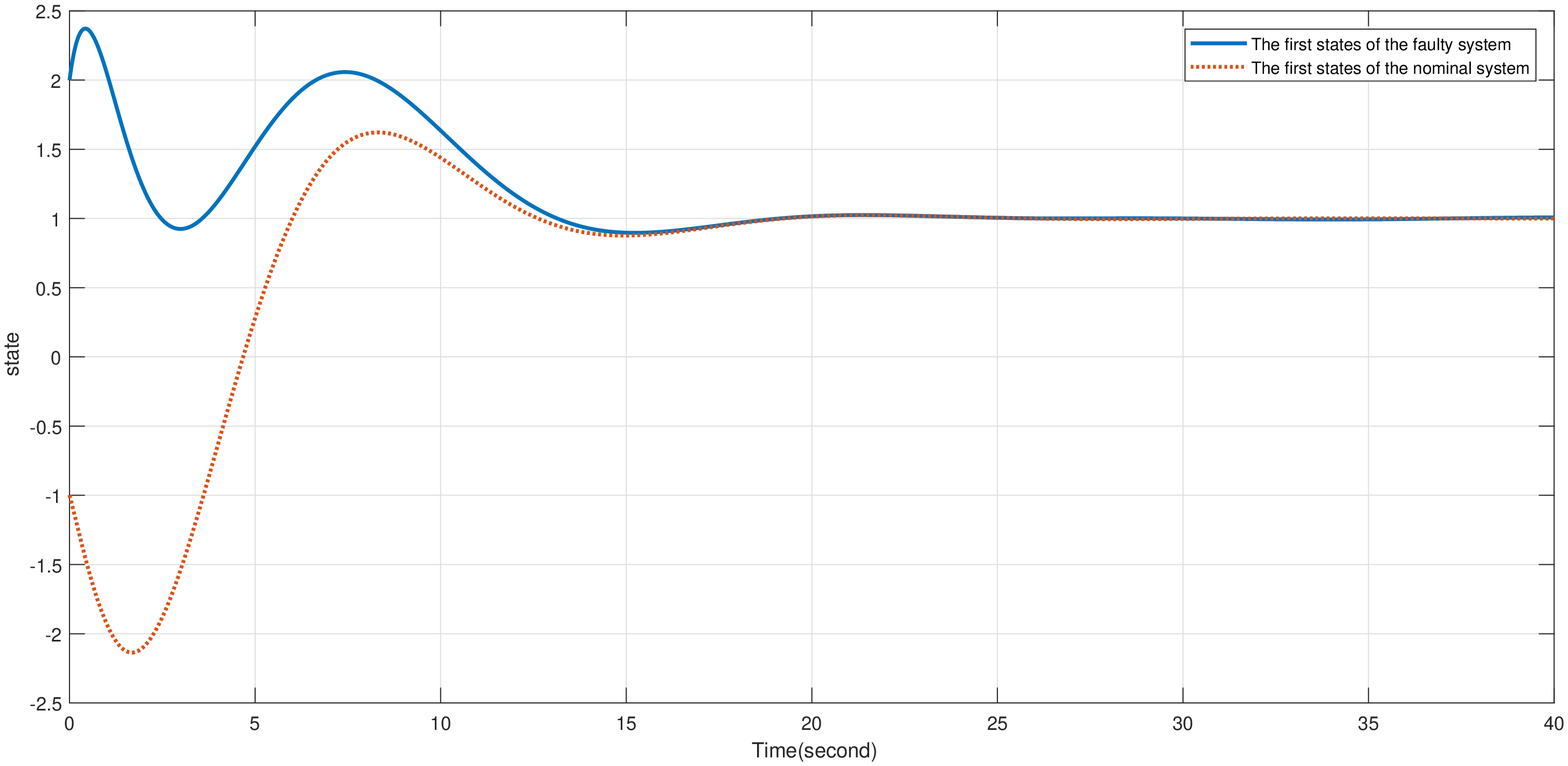}
	\caption{The first states of the faulty system and nominal system.}
	\label{Blockdiagram}
\end{figure}

\begin{figure}[!ht]
	\centering
	\includegraphics[scale=0.25]{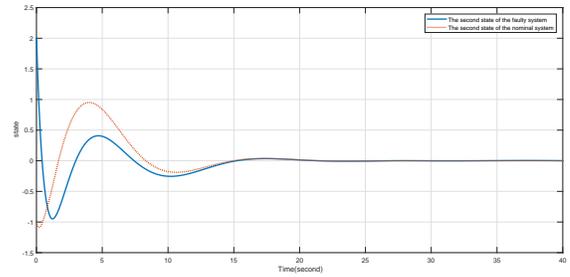}
	\caption{The second states of the faulty system and nominal system.}
	\label{Blockdiagram}
\end{figure}

\begin{figure}[!ht]
	\centering
	\includegraphics[scale=0.25]{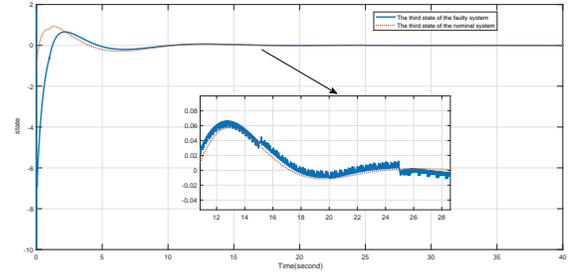}
	\caption{The third states of the faulty system and nominal system.}
	\label{Blockdiagram}
\end{figure}

Also, Figure 7 is zoomed in to show the fault, disturbance and time-varying additive actuator fault occurrence.

\begin{figure}[!ht]
	\centering
	\includegraphics[scale=0.25]{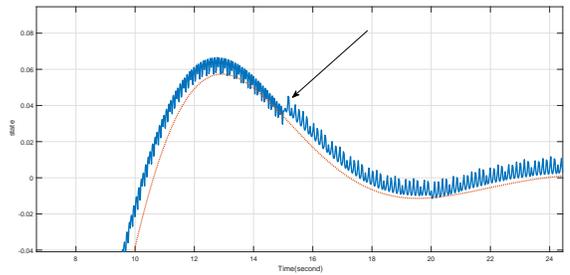}
	\caption{Zoom in when the fault occurs in 15 second.}
	\label{Blockdiagram}
\end{figure}

\begin{figure}[!ht]
	\centering
	\includegraphics[scale=0.25]{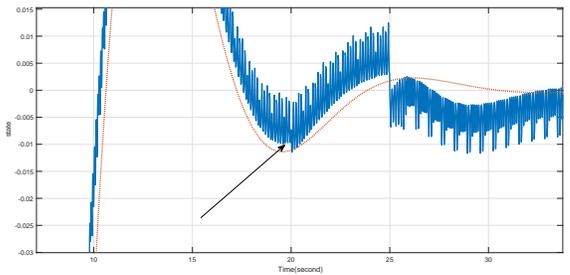}
	\caption{Zoom in when the disturbance occurs in 20 seconds.}
	\label{Blockdiagram}
\end{figure}

\begin{figure}[!ht]
	\centering
	\includegraphics[scale=0.25]{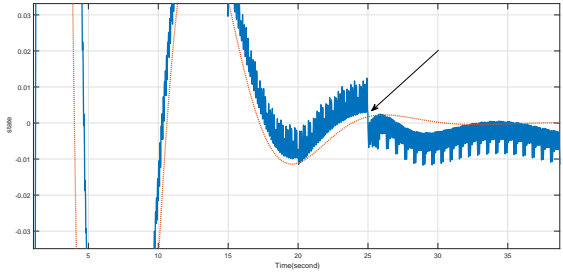}
	\caption{Zoom in when the time-varying actuator fault occurs in 25 seconds.}
	\label{Blockdiagram}
\end{figure}
Fig.11 and Fig.12 show the control input in the nominal mode and the faulty mode.
\begin{figure}[!ht]
	\centering
	\includegraphics[scale=0.25]{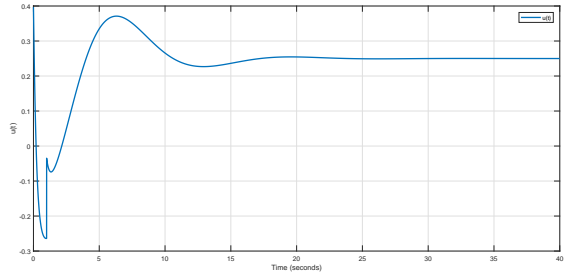}
	\caption{The control input in the nominal mode.}
	\label{Blockdiagram}
\end{figure}

\begin{figure}[!ht]
	\centering
	\includegraphics[scale=0.25]{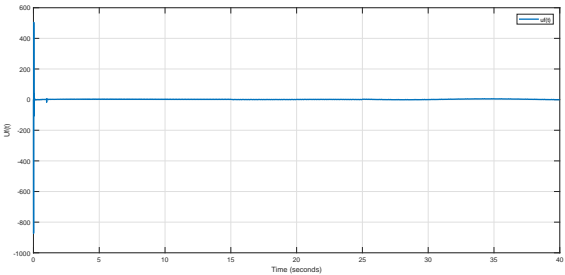}
	\caption{The control input in the faulty mode.}
	\label{Blockdiagram}
\end{figure}

Also, the adaptive Controller Coefficients are shown in Fig. 11, Fig. 12 and Fig. 13.
\begin{figure}[!ht]
	\centering
	\includegraphics[scale=0.25]{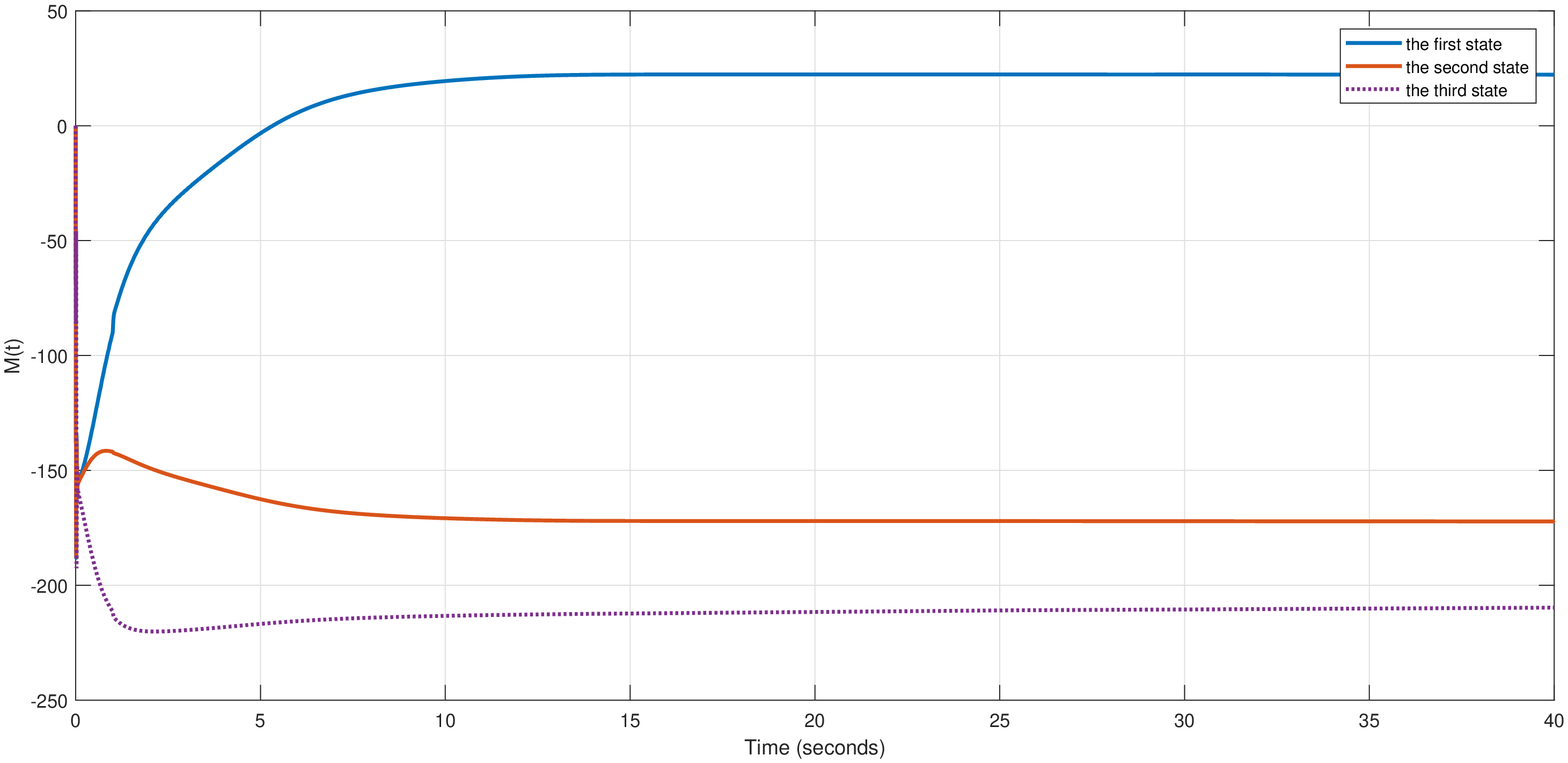}
	\caption{Update parameter of ${M}(t)$.}
	\label{Blockdiagram}
\end{figure}

\begin{figure}[!ht]
	\centering
	\includegraphics[scale=0.25]{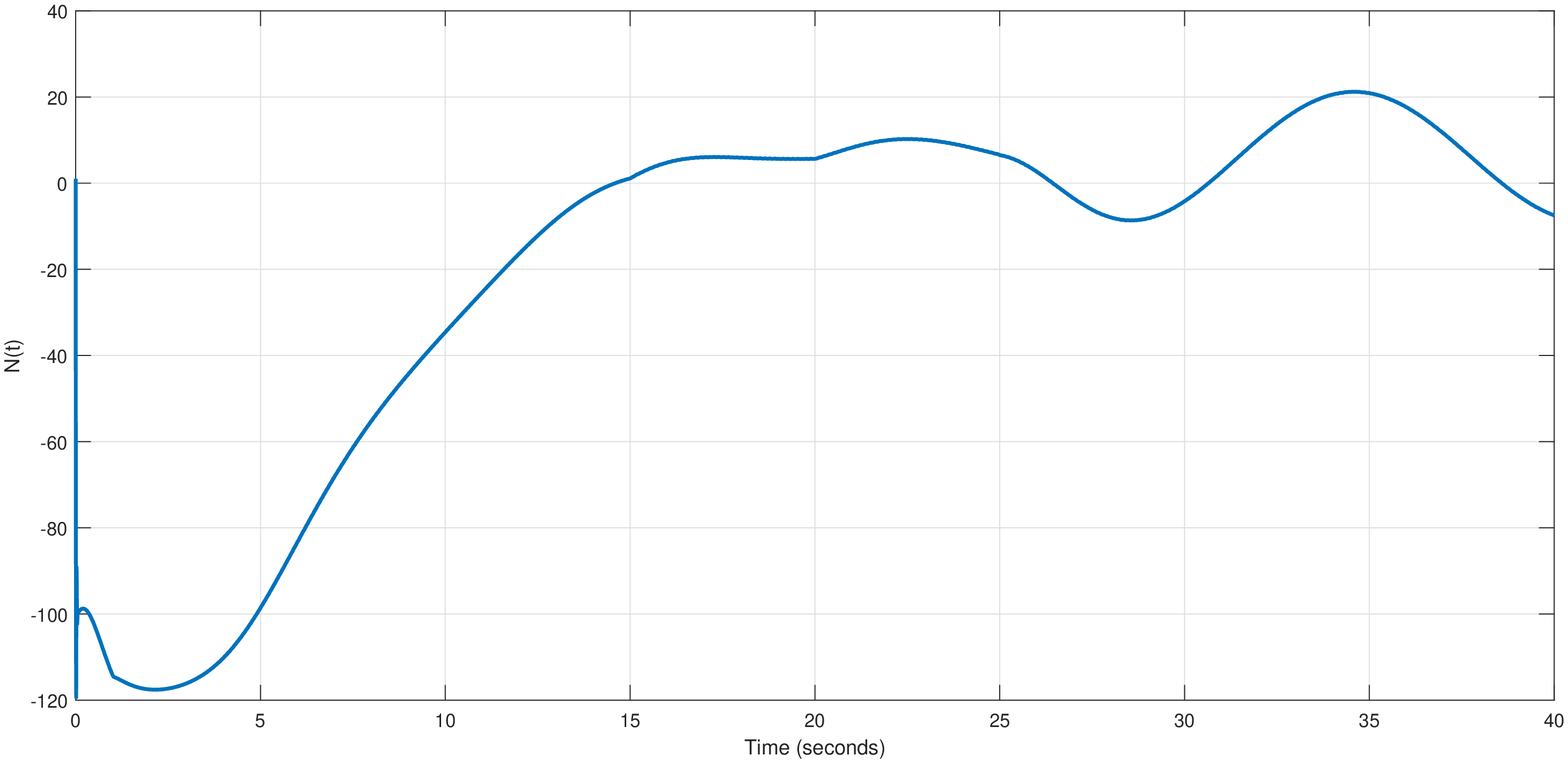}
	\caption{Update parameter of ${N}(t)$.}
	\label{Blockdiagram}
\end{figure}

\begin{figure}[!ht]
	\centering
	\includegraphics[scale=0.25]{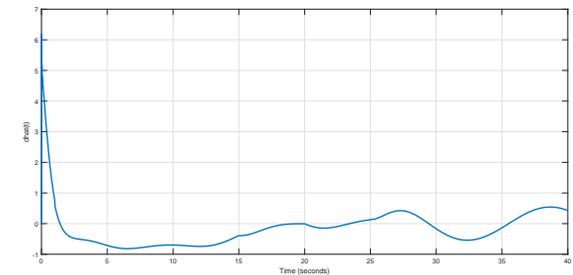}
	\caption{Update parameter of ${\hat{{d}_{t}}}(t)$.}
	\label{Blockdiagram}
\end{figure}
\section{Conclusion}
In this paper, An Adaptive Fault-tolerant Controller has been considered for a class of the affine nonlinear system with nonlinear input gain. This is based on fault-hiding and does not require a separate FDI unit. Also, the design of virtual actuator is based on the feasible solution of LMI. The convergence of  faulty and  the desire system is guaranteed. The faults affect and external disturbances are hidden and furthermore, the system is robust against faults. Finally, the simulation result shows the promising performance of the proposed methodology.


	\end{document}